\documentstyle[aps,preprint,prl]{revtex}
\begin{document}
\draft
\title{Reconstruction Mechanism of fcc Transition-Metal (001) Surfaces}
\author{Vincenzo Fiorentini,$^{(1)}$\cite{vif_ad} Michael Methfessel,$^{(2)}$
         and Matthias Scheffler$^{(1)}$}
\address{$^{(1)}$ Fritz-Haber-Institut der Max-Planck-Gesellschaft,
	 Faradayweg 4-6, D-14195 Berlin, Germany \protect \\
$^{(2)}$   Institut f\"ur Halbleiterphysik, Walter-Korsing-Strasse 2,
D-15230 Frankfurt/Oder, Germany}
\date{Received 23 May 1993}

\maketitle
\begin{abstract}
The reconstruction mechanism of  (001) fcc transition metal surfaces is
investigated using a full-potential all-electron electronic structure
method together with density-functional theory.
 Total-energy calculations  confirm the
experimental finding that a close-packed quasi-hexagonal overlayer
reconstruction is possible for the   late 5$d$-metals Ir, Pt, and Au,
while it is disfavoured in the isovalent 4$d$ metals (Rh, Pd, Ag).
The reconstructive behaviour is driven by the tensile surface stress
of the unreconstructed surfaces; the stress is significantly
larger in the 5$d$ metals than in 4$d$ ones, and only in the former case it
overcomes the substrate resistance to the required geometric
rearrangement. It is shown that the surface stress for these systems
 is due to $d$ charge depletion from the surface layer,  and that
 the cause of the 4th-to-5th row stress difference is the importance
of  relativistic effects in the 5$d$ series.
\end{abstract}

\pacs{PACS numbers : 68.35.-p, 68.35.Bs, 68.35.Md}

The (001) surfaces of some fcc transition and noble metals are known to
reconstruct to a   close-packed quasi-hexagonal (hex) overlayer arrangement,
periodically matching   the (001)  square substrate \cite{rec}. Due to the
different symmetries of  overlayer and substrate,
commensurate hex geometries   have rather long periodicities,
typical observed surface cells being 5$\times$1 and 20$\times$5.
The phenomenon presents a number of  interesting aspects. First, the
reconstruction  is seen only  at the  end of the 5$d$ transition series,
for the metals Ir, Pt, and Au, while their 4$d$
isoelectronic upper neighbours Rh, Pd, and Ag do not  reconstruct. Second,
the observed reconstructions of Ir, Pt, and Au, though not identical,
 are qualitatively very  similar; in view of  the differences in the
electronic structures of those materials,  this strongly suggests
that surface electronic-structure details are not of primary importance.
Third, it is known \cite{rec} that the (001) surfaces of late 5$d$ metals can
be rather easily forced to switch between  the  unreconstructed and the
reconstructed phase by deposition or removal of small amounts of adsorbates,
which indicates that the energy difference between the two phases is small.

The aim of this Letter is to study this class of reconstructions
by a density-functional-theory  treatment
\cite{gd,fp,tm}, and to understand
the underlying physical mechanism  establishing a simple model picture
of the energetics of the phenomenon.
In the following, we first present  a direct  calculation of the heat of
reconstruction of the 1$\times$1 to 5$\times$1 reconstructive
transition for the (001) surfaces of the isoelectronic pair
Pd and Pt, showing that reconstruction is disfavoured in the former (4$d$) and
permitted  in the  latter (5$d$), in accordance with experiment.
To clarify the reconstruction mechanism, we calculate the surface energy and
stress for the relevant {\it unreconstructed} surfaces, showing
that they are subjected to a large tensile stress -- that is, they
tend to prefer a smaller in-plane lattice constant and a higher in-plane
atomic density. While indeed achieving close-packing of the surface
layer, the quasi-hexagonal reconstruction also causes a modification of the
surface-substrate bonding topology, effectively reducing the average atomic
coordination; we demonstrate
that  {\it (a)} the reconstruction is determined by the balance
between the energy gain associated with the increase in atomic density
at the surface, and the energy lost upon reconstruction
due to the disruption or stretching of bonds
between mismatched top and subsurface layers,
and that {\it (b)} only in 5$d$ metals the surface energy gain due to
surface density increase is large enough as to overweigh the mismatch energy
loss, thus making the reconstruction favourable in those  elements only.
This is due to the fact that the  surfaces
of end-of-series 5$d$-metals (Ir, Pt, Au)
are subiect to  twice as large a tensile stress than
those of the isovalent 4$d$ metals (Rh, Pd, Ag),
 and thus gain much more energy upon close-packing.
We then discuss the origin of surface stress, which is due to
depletion of $d$ charge from the surface, and  the  relativistic effects
responsible for the enhancement of this mechanism in the 5th row.

\paragraph*{Technical matters --}
The  calculations  were performed using
density functional theory (DFT) together with the  local density
approximation (LDA)\cite{gd} and the all-electron full-potential LMTO
method\cite{fp}. The
unreconstructed  1$\times$1 surface is simulated by 7-layer slabs, separated
by about 10 layers of vacuum, the {\bf k}-point summations being done on a
15-point mesh in the irreducible part of the surface Brillouin zone  (ISBZ)
of the 1$\times$1 surfae cell.
For the $5\times 1$ surface, the summation was performed
on a  somewhat denser grid (32-points in the 5$\times$1 ISBZ), and
a 5-layer slab was
used. Details on the method and on its  previous applications to surfaces
can  be found in Ref.\cite{tm}, while a full report of the present calculations
will be given elsewhere\cite{long}.

Part of the calculations described in the following are basically concerned
with the  elastic response of the surface.
 The appropriate reference system is therefore
the stress-free crystal at theoretical equilibrium. It is worth  pointing out
a general problem of DFT-LDA calculations for transition
metals. Scalar relativistic calculations give excellent results
for the equilibrium properties of bulk 5$d$ metals, and such
treatment appears indeed to be vital for a proper description of these systems;
in 4$d$ metals, on the other hand, the relativistic treatment
gives   lattice constants about 2--2.5\% too small, and as a consequence it
produces considerable errors in the theoretical bulk moduli (about 30 to 40\%
too large),
while non-relativistic calculations give excellent equilibrium crystal
properties  (in 3$d$ metals the scalar-relativistic treatment gives
 even worse results). Although these errors may be considered acceptable
(they are at  the commonly accepted  limit of LDA-standard accuracy), one is
nevertheless confronted with the fact that, for the elements in
question, a relativistic treatment  may induce  spurious effects in
the elastic response, deriving
from the incorrect equilibrium crystal ground state it produces.
We therefore chose to present here non-relativistic results for the 4$d$
metals. Accurate tests, to be discussed elsewhere\cite{long}, show indeed
that the picture presented here remains unaltered if a scalar
relativistic treatment is adopted for the 4$d$ metals.

\paragraph*{5$\times$1 reconstruction --}
For the  computational study  of this prototypical
hex-overlayer system, we assumed the structure inferred from LEED
data\cite{5x1}, in which 6 surface atoms are packed together
on top of each 5 substrate  atoms in a (11) direction on the (001) surface,
as sketched in Fig.\ref{hex}.
To obtain a safer comparison,  we also calculated the {\it unreconstructed}
5$\times$1 surface ({\it i.e.}, a 1$\times$1 cell repeated 5 times)
with the same technical ingredients, obtaining a  surface energy
within 2\% of that calculated using the 1$\times$1 surface cell.
Since the reconstructed geometry was not relaxed,
in view of typical energy changes upon relaxation\cite{tm}, we estimate
the  overall error bar of the calculation to be in the  order of 0.05 eV per
$1\times 1$ area (or 6 meV/\AA$^2$ for Pd and Pt).

The heat of reconstruction $E_r$ is defined to be the difference of the total
energies of the unreconstructed slab plus two bulk atoms and of the
reconstructed slab, divided by ten to refer the energy to the area of the
1$\times$1 (001) surface cell. A positive value thus indicates
the reconstruction to be energetically favoured.
The numerical result is $E_r = -0.21$ eV in Pd and  $E_r = -0.03$
eV in Pt, indicating that Pd will not
reconstruct, whereas in Pt the reconstructed and unreconstructed phases
have equal energies within the accuracy of the calculation.
This result agrees with the fact that both
phases are observed for Pt depending on the experimental
conditions, while the reconstruction has not been observed in Pd.

Since the reconstruction increases the surface atomic density,
atoms have to be added to the surface layer. In equilibrium conditions, it
can be assumed that the additional atoms come from the bulk, which actually
means that they come from kink sites at surface steps.
Indeed, the atom chemical potential in thermal equilibrium equals the
 the crystal cohesive energy. We note in passing
that if the surface were coupled, in non-equilibrium conditions,
to a reservoir with a lower chemical potential\cite{mic}, the energy cost per
additional  atom would be lower, and the heat of reconstruction would
increase ({\it i.e.} the reconstruction would be more favoured).

\paragraph*{Surface stress vs. surface-substrate mismatch --}
To gain insight into the mechanism driving the reconstruction,  we calculated
the surface energies, stresses and relaxations for  the {\it unreconstructed}
(001) surfaces of the relevant fcc metals. From calculated total energies of
slabs and bulk at various in-plane lattice constants, we obtain
the   strain derivative of the surface energy,
$\tau =  d\, \sigma/d\, \epsilon$. Due to the fourfold symmetry of
the system, $\tau$ is isotropic and can be expressed as the derivative
$\tau =  d\, \sigma/d\, a$ with respect to the area $a = A/A_0$ normalized
to the equilibrium area $A_0$  of the unreconstructed  1$\times$1
surface cell.
We name $\tau$ the  {\it excess} surface stress, as it gives a quantitative
measures of the {\it change} in surface energy,
$\sigma(A) - \sigma(A_0) =  \tau\,\delta
a$, which would result from a relative area variation $\delta a =
(A-A_0)/A_0$.
We note that, in a related context,
use has been made in the past of the total surface
stress\cite{needs} of the unreconstructed surface, which is
the sum of $\tau$ and of the surface energy $\sigma$ at zero strain. Though,
the relevant quantity here is indeed the excess part of the stress, since,
while $\sigma$ itself is a fixed cost of formation for the unreconstructed
surface
at  the in-plane lattice constant  determined by the underlying bulk, the
value and sign of $\tau$  indicate the tendency
to reduce the  surface energy by changing the surface  atomic density.
In particular, a positive (or tensile) excess  stress indicates
that the surface tends to contract and to attain a higher atomic density.
Of course, as discussed below, the bonds to the substrate will strongly
counteract the surface layer tendency to change its in-plane lattice constant
A further energy cost originates, as seen above, from the need to
increase  the atomic  density  of the surface layer.

Our  results are presented in Table I. The excess
surface stress is tensile in all cases, and it increases considerably
going from the 4th to the 5th row.
We now show that the large  tensile excess surface
stress is the driving force of the
quasi-hexagonal reconstruction of the Ir, Pt, and Au (001) surfaces.
To attain close-packing, the surface
layer has to rearrange to a different geometry and must thereby overcome the
energy cost of bond rearrangement; we estimate this cost  by
ideally splitting the heat of reconstruction into a ``gain" and a ``loss";
the latter, the bond rearrangement contribution $\Delta E_b$
to the heat of reconstruction $E_r$, is obtained by subtracting from $E_r$
the stress-related surface energy gain $\delta \sigma = -\tau\,\delta a$
obtained by a reduction of the surface area per atom to the value of the
(111)  surface ({\it i.e.} about 14\%; the sign of $\delta \sigma$ is
chosen consistently with our convention for the heat of reconstruction).

The values for the 4th and 5th row are rather close: $\Delta E_b = -0.36$ eV
for Pd, and $\Delta E_b = -0.41$ eV for Pt, referred to 1$\times$1 area.
We conclude that the favourable balance for the reconstruction in 5$d$ metals
as opposed to 4$d$ metals
is indeed determined by an exceptionally large stress-related energy gain
in the former case, driving the reconstruction against a bond rearranging cost
which is about the same in both series.
Inspection of the reconstructed geometry reveals a
 reduction in number of bonds between
surface and first substrate layer. Thus, the bond
rearrangement cost is mostly due to  the
{\it average} coordination of subsurface and surface atoms
being decreased, despite the increase of  in-plane coordination
in the reconstructed top   layer.

It is worth noticing that the balance between surface contraction and
surface-substrate mismatch is favourable in 5$d$ metals
and unfavourable in 4$d$ metals
because of the {\it magnitude} of the surface excess stress; the
mere {\it sign} of the latter is in itself not sufficient to decide whether or
not the reconstruction will actually take place.
Further, we emphasize that even  the magnitude of the stress is meaningful
as a reconstruction predictor only when related to the substrate bonding
resistance to the reconstruction. This is expecially relevant when considering
systems in different positions in the transition series:
for instance, Rh (001) has a larger stress than Au (001),
but the $d$ bonding to the substrate is very much stronger in the former case,
so that indeed Rh (001) is not able to reconstruct, while Au is.

\paragraph*{Origin of surface stress  enhancement --}
For $d$ metals, an
explanation of the  surface stress and of its magnitude can be found in the
competition between $sp$ and $d$ bonding, and how the balance
is modified at the surface. The accepted description of  bonding
 for a transition-metal series\cite{friedel}
is that as the $d$ occupation $n_d$ increases from 0 to 10,
bonding, non-bonding, antibonding states are successively filled. Neglecting
$sp$ electrons, this leads to a parabolic bond strength as function of
$n_d$, with a maximum around $n_d \simeq 5$.
Including   $sp$ electrons in the picture, it is found that
throughout most of the series the $d$ electrons form localized bonds
which tend to contract the crystal, while the more diffuse $sp$ electrons
exert an outwards pressure\cite{pett}.
This balance is reversed when the $d$ band is nearly full; the $sp$
electrons now bind the crystal, while the full  $d$ shell
tends to resist lattice contraction.

As an indication of how the balance in the bulk is perturbed bu
the surface, we inspect the $sp$ and $d$ charges at the surface and in
the bulk. A Mulliken population analysis\cite{mull} shows
that the total charge $Q_S$ at the surface is smaller than the charge
$Q_B$ in the bulk, {\it i.e.} $\delta Q = Q_S - Q_B < 0$. More specifically,
this depletion is mostly of $d$ character
($\delta Q^{\it d} = Q_S^{\it d} - Q_B^{\it d} < 0$), while
the $sp$ charge has slight net increase
($\delta Q^{\it sp} = Q_S^{\it sp} - Q_B^{\it sp} \geq 0$).
The total layer charge as a function of position into the slab shows
first a depletion (mostly $d$),
 then an enhancement in the first sublayer (mostly $d$),
and finally it rapidly saturates into the bulk.
As seen from Table I, the magnitude of the stress correlates with the
amount of charge depletion from the surface layer.
Since the depleted charge is predominantly of $d$ character, the
tensile surface stress in the late transition metals ($n_d > 5$) can be
explained  as a consequence of
de-occupying antibonding $d$ states at the high-energy end of the
surface density of states (DOS).
Thereby, the bond strength between surface atoms is increased.
Direct inspection of the charge density shows that $d_{xy}$ orbitals are
indeed depleted with respect to bulk occupation; the  depletion takes place by
partial charge transfer to other $sp$-hybridized $d$ states\cite{long}.
The mechanism giving rise to surface stress is an ``internal
conversion'' in the $d$ shell assisted and enhanced by hybridization with $sp$
orbitals\cite{long}, and is thus quite different from that causing
the stress of nearly-free-electron metal surfaces\cite{needs}.
In particular, due to the key role of $d$ states,
jellium-based descriptions might be inapplicable to  the
present cases.

Given the above findings,
 the sizably larger surface stress in the 5th-row metals is
easy to understand: the origin are the strong relativistic
effects in these systems. It is known that
the lattice constants  of Ir, Pt and Au are about
the same as those of Rh, Pd, and Ag despite the much larger atomic
size of the 5$d$ elements, and the 5$d$  bulk moduli are
 larger than those of the 4$d$'s by a factor of about 1.5--2.
This is a consequence of the  enhanced bonding in 5th-row  metals as
compared to 4$d$ metals, due to relativistic effects\cite{pikko}.
 The $6s$ and $6p$ electrons contract and  lower their energy due to the
mass-velocity term, so  that $sp$
occupation is increased. As a consequence antibonding $d$ states are
emptied,  thus the bonding is overall  enhanced. In terms of the DOS, this
increased bonding  $sp$-$d$ hybridization in a 5$d$ metal corresponds to a
longer $sp$ tail, and to a $d$-band complex which is wider and closer in
energy to the Fermi level than in a 4$d$ one. Hence in 5$d$ metals
the narrowing and upward  shift of the DOS at the surface will produce
a larger $d$ depletion (partly in favour of $sp$ charge).
A direct demonstration of the effect is provided by a non-relativistic
calculation for Pt. The lattice constant is 5\% larger than experimentally
observed, and the bulk modulus is only 60 \% of its actual value, {\it i.e.}
close to that of Pd. Most importantly, the surface stress is indeed
 found to decrease by a factor of 2
and the surface energy by 30\%, while the bulk-to-surface charge
differences are also significantly smaller:  overall the surface quantities
for non-relativistic Pt resemble rather closely those of Pd\cite{long}.
The strength of relativistic effects thus appears to be the relevant
difference  between 4th and 5th-row metals in this context, and it may be
identified as the ultimate cause for the reconstruction of 5th-row fcc (001)
surfaces.

\paragraph*{Summary --} We have calculated the quasi-hexagonal reconstruction
of the (001) surfaces of representative fcc  transition metals, finding it to
be  favoured for the late
5$d$ metals and not for 4$d$ metals, in accordance with experiment.
A correlation has been established between reconstruction and magnitude of
the surface stress calculated {\it ab initio} for
the unreconstructed surfaces: the reconstruction results from a delicate
balance between surface-substrate mismatch and stress-related energy gain.
Only in the case of 5$d$ metals is the  latter gain
 large enough to actually drive the reconstruction against the
substrate resistance to misregistry, which is comparable for isoelectronic
 systems ({\it e.g.} Pd and Pt).
The origin of the surface  stress is the
 $d$ charge  depletion at the surface, caused by enhanced $sp$ hybridization;
 the remarkable stress enhancement in  5$d$ metals is due to the major
 relativistic effects on the 6$s$ and 6$p$ shells.
%
%

%
\begin{table}
\begin{tabular}{l|rrrrrrr}
\multicolumn{1}{c}{ } &
\multicolumn{1}{c}{ $\sigma$} &
\multicolumn{1}{c}{ $\tau$} &  
\multicolumn{1}{r}{ $\Delta d_{12}$} &
\multicolumn{1}{c}{ W } &
\multicolumn{1}{c}{ $\delta Q$ } &
\multicolumn{1}{r}{ $\delta Q^{sp}$ } &
\multicolumn{1}{r}{ $\delta Q^d$ } \\
\tableline
Rh   & 1.26 & 1.94 & --4.5 &  5.24 & --0.30 & 0.08 & --0.38 \\
Pd   & 0.91 & 1.05 & --0.8 &  5.30 & --0.19 & --0.08 & --0.11 \\
Ag   & 0.59 & 0.88 &  --1.9 &  4.43 & --0.11 & 0.02 & --0.13 \\
\tableline
Ir   & 1.73 & 2.94 & --3.0 &  5.92 & --0.42 & 0.24 & --0.66 \\
Pt   & 1.21 & 2.69 &   0.0 &  6.11 & --0.36 & 0.15 & --0.51 \\
Au   & 0.75 & 1.62 & --1.0 &  5.61 & --0.20 & 0.10 & --0.30 \\
\end{tabular}
\caption{Surface energy and stress (in eV/(1$\times$1) cell area),
top-layer  relaxation (percentage of interlayer spacing), and work
function (eV) of  the (001) surfaces of 4$d$ and 5$d$ fcc transition and
noble metals. The variation of total, $sp$ and $d$ charge between bulk and
surface (electrons) obtained by Mulliken analysis is given.}
\end{table}
\begin{figure}
\caption{Side and top view of the 5$\times$1 reconstruction.}
\label{hex}
\end{figure}
\end{document}